\documentclass[prl,twocolumn,superscriptaddress,preprintnumbers,%
showpacs,amsmath,amssymb]{revtex4}

\usepackage{graphicx}

\begin{document}

\title{Order parameters with higher dimensionful composite fields}
\author{Yoshiki Watanabe}
\email{watanabe@nt.phys.s.u-tokyo.ac.jp}
\affiliation{Department of Physics, University of Tokyo,
 7-3-1 Hongo, Bunkyo-ku, Tokyo 113-0033, Japan}
\author{Kenji Fukushima}
\email{kenji@lns.mit.edu}
\affiliation{Department of Physics, University of Tokyo,
 7-3-1 Hongo, Bunkyo-ku, Tokyo 113-0033, Japan}
\affiliation{Center for Theoretical Physics, Massachusetts Institute
 of Technology, Cambridge, Massachusetts 02139}
\author{Tetsuo Hatsuda}
\email{hatsuda@nt.phys.s.u-tokyo.ac.jp}
\affiliation{Department of Physics, University of Tokyo,
 7-3-1 Hongo, Bunkyo-ku, Tokyo 113-0033, Japan}

\begin{abstract}
We discuss the possibility of the spontaneous symmetry breaking
characterized by order parameters with higher dimensionful composite
fields. By analyzing general Ginzburg-Landau potential for a complex
scalar field $\phi=\phi_1+\mathrm{i}\phi_2$ with $O(2)$ symmetry, we
demonstrate that a phase characterized by
$\langle\phi_1^2-\phi_2^2\rangle\neq0$ with 
$\langle\phi_1\rangle=\langle\phi_2\rangle=0$ is realized in a certain
parameter region. To clarify the driving force to favor this  phase,
we study the $O(2)$ $\phi^6$ theory in three dimensions. 
\end{abstract}
\preprint{MIT-CTP 3457}
\maketitle


The spontaneous symmetry breaking (SSB) should be enumerated as one of
the key concepts in modern physics, in particular in condensed matter
and elementary particle physics \cite{nam61}. Even if some global
symmetry is manifest in the Hamiltonian, the ground state and excited
spectra do not have to reflect the symmetry manifestly. The order
parameter, which is a measure of SSB, is defined by the ground state
expectation value of an operator being \textit{variant} under the
symmetry transformation. The choice of the operator is constrained by
the symmetry breaking pattern. The purpose of this Letter is to draw
attention to the issue of SSB through such order parameters with
higher canonical dimensions. As we will argue, unbroken discrete
groups generally require operators with higher canonical dimensions.

Let us first give a summary to construct generic order parameters.
Suppose $X$ and $Q_a$ be a symmetry group and the generators of the
symmetry transformation, respectively. We take an operator
$\mathcal{O}$ which transforms non-trivially under $Q_a$;
$[\mathrm{i}Q_a,\mathcal{O}]\neq0$. SSB takes place if we have such an
$\mathcal{O}$ that $\langle0|[\mathrm{i}Q_b,\mathcal{O}]|0\rangle\neq0$
for some $b$
 \footnote{Precisely speaking,
 $\langle0|[\mathrm{i}Q_b,\mathcal{O}]|0\rangle$ is ill-defined as it
 is when SSB occurs. Nevertheless,
 $\int\mathrm{d}x\langle0|[\mathrm{i}j_b,\mathcal{O}]|0\rangle$ with
 $Q_b=\int\mathrm{d}x\,j_b$ can be finite and make sense.}.
If $Q_b$ is a generator of $X$ but is outside of its subgroup $Y$, the
symmetry is spontaneously broken from $X$ to $Y$ with an order
parameter $\langle0|[\mathrm{i}Q_b,\mathcal{O}]|0\rangle$.
$\mathcal{O}$ can be chosen to be any operators as long as it is
invariant under $Y$. 

In particular, if $Y$ contains a discrete
subgroup of $X$, SSB generally requires higher dimensionful composite
operators for $\mathcal{O}$ because the elementary operators are not
invariant in $Y$.
We pursue such exotic realization of SSB not only for academic
interest but for phenomenological importance. In Quantum
Chromodynamics (QCD) we sometimes encounter possibilities of higher
dimensionful operators. A well-known example is the four-quark
operator introduced for a non-standard chiral symmetry breaking
pattern in $N_{\text{f}}$-flavors \cite{ste98} where
$X=SU_{\text{L}}(N_{\text{f}})\times SU_{\text{R}}(N_{\text{f}})$ with
$Y=SU_{\text{V}}(N_{\text{f}})\times Z_{\text{A}}(N_{\text{f}})$.
Although this is proven to be forbidden in QCD at zero and finite
temperature, it may be realized at finite baryon chemical potential.
Interestingly enough, the remaining $Z_{\text{A}}(N_{\text{f}})$
symmetry could change the hadron spectrum significantly \cite{gir01}.
Exotic hadron spectrum in a similar nature\cite{son00} is indeed known
to be realized in the color-flavor-locking (CFL) phase in color
superconductivity where $U_{\text{B}}(1)$ is broken down to $Z(2)$
subgroup \cite{alf99}. In particular, the existence of light kaonic
excitations associated with this SSB opens a possibility for the kaon
condensation in the CFL phase \cite{bed02}. Therefore, in connection
with these phenomenological concerns, it is an important issue to
probe SSB with an unbroken discrete subgroup and with higher
dimensionful order parameter from the generic point of view.

In order to look into the problem in a concrete setting, let us take
an $O(2)$ symmetric model with one complex scalar field
$\phi= \phi_1 + \mathrm{i} \phi_2$. If we take
$\mathcal{O}=\mathrm{Im}\phi=\phi_2$ which is not symmetric in any of
the $O(2)$ transformation, we have
$\delta\phi_2 \propto \mathrm{Re}\phi=\phi_1$. It follows that the
order parameter for the symmetry breaking pattern
$X=O(2) \to \text{none}$ is
$\langle\phi_1\rangle \equiv \langle0|\phi_1|0\rangle$. This is,
however, not a unique symmetry breaking pattern possible. Consider
$X=O(2)$ and $Y=Z_2$ with $Z_2$ being a discrete subgroup with the
operation; $\phi_{1,2} \to -\phi_{1,2}$. In this case
$\mathcal{O}=\mathrm{Im}\phi^2=2\phi_1\phi_2$, which is $Y$ invariant,
leads to an order parameter,
\begin{equation}
 \mathrm{Re}\langle\phi^2 \rangle = \langle\phi_1^2-\phi_2^2 \rangle.
\label{eq:2nd}
\end{equation}
Note that the difference between $\langle\phi_1^2\rangle$ and
$\langle\phi_2^2\rangle$ is an appropriate order parameter 
 instead of 
the individuals because the common constant cancels out in the former.
One can construct $k$-th composite order parameters for $O(2) \to Z_k$
in the same way;
\begin{equation}
 \mathrm{Re}  \langle \phi^k \rangle.
\end{equation}
If $\mathrm{Re} \langle\phi\rangle$ has a non-vanishing value, the
$O(2)$ symmetry is totally broken so that the higher dimensionful
order parameters, $\mathrm{Re} \langle\phi^k \rangle$, are non-zero
for all $k\ge 1$. The SSB with $O(2) \to Z_k$ can be characterized by
the condition;
$\mathrm{Re} \langle\phi\rangle = \cdots =
 \mathrm{Re} \langle\phi^{k-1} \rangle =0$ and yet
$\mathrm{Re} \langle\phi^k \rangle \neq 0$.

Application of the above argument to $O(N)$ symmetric models with
$N>2$ is straightforward. For the SSB pattern,
$ O(N) \to O(2) \to Z_k$, the higher dimensionful order parameter can
be defined in the same way as above with respect to the $O(2)$
subgroup.
 

Now we shall go on to investigate whether there is a situation where
the above SSB pattern, $O(2) \to Z_k$, is indeed realized. We will
focus on a simplest case, $k=2$, namely the phase characterized by
$\langle\phi_1^2-\phi_2^2\rangle$. For later purpose, we introduce
complex variables $W$ and $H$, and a real variable $F$ as follows,
\begin{equation}
 \begin{split}
 & W=\langle \phi^2 \rangle  = \langle\phi_1^2-\phi_2^2\rangle
   +2\mathrm{i}\langle\phi_1\phi_2\rangle,\\
 & F=\langle \phi^{\dagger} \phi \rangle
   =\langle\phi_1^2+\phi_2^2\rangle,\\
 & H=\langle \phi \rangle = \langle\phi_1 +\mathrm{i} \phi_2\rangle.
 \end{split}
\label{eq:WFH}
\end{equation}

To clarify the phase structure in the $W$-$F$-$H$ space, let us
consider a Ginzburg-Landau type effective potential with $O(2)$
symmetry up to terms of $\phi^6$ order which are at least necessary to
induce a phase with $W\neq 0$;
\begin{align}
 & V_{\text{eff}}[W,F,H] \notag\\
 & = (d_1+d_2 F+d_3 F^2)|H|^2+(d_4+d_5 F)|H|^4+d_6|H|^6 \notag\\
 &\qquad +(c_1+c_2 F)(W^{\dagger}H^2  +H^{\dagger 2}W) \notag\\
 &\qquad +a_1 F+a_2 F^2+a_3 F^3+(b_1+b_2 F)|W|^2.
\label{eq:general}
\end{align}
The coefficients $a_i$, $b_i$, $c_i$, and $d_i$ are considered to be
arbitrary coupling constants at the present stage. Taking a specific
model with $O(2)$ symmetry, one may determine their magnitudes and
relations. For more than two dimensions, one needs to subtract the
short distant singularities in $F$ and $W$ to make them finite. This
will be discussed in more detail later and we assume a simple cutoff
in the short distant part here.
  
The standard phase corresponding to the SSB pattern,
$O(2) \to \text{none}$, is characterized by $H\neq 0$ (and thus
$W\neq 0$), while the non-trivial phase corresponds to $W\neq 0$ with
$H=0$. Note that $F\neq 0$ in both cases. For convenience, we call the
latter phase as ``the WFH phase'' after the notation in
Eq.~(\ref{eq:WFH}). To find the condition for the WFH phase, we choose
$d_i$ appropriately so as to realize $H=0$. Then the GL potential is
reduced to only the terms containing $a_i$ and $b_i$ in
Eq.~(\ref{eq:general}). Among these coupling constants, we can
eliminate two of them by rescaling $F$ (and $W$) and coupling
constants. Then one finds
\begin{equation}
 V_{\text{eff}}[W,F] = a_1' F\pm F^2+F^3+(b_1'+b_2' F)W^2,
\label{eq:general_fw}
\end{equation}
where $W$ is chosen to be real and positive without loss of generality
and as a consequence we have a constraint, $F\ge W \ge 0$. Note that
the sign of the $F^3$ term should be positive to guarantee the
stability of the potential, while it is not necessary for the
coefficient of the $F^2$ term.
 
An instability toward $W\neq0$ is induced by the negative $W^2$ term
in Eq.~(\ref{eq:general_fw}). Therefore, once the instability to the
WFH phase occurs, $W=F$ is realized because of the constraint
$F\ge W \ge 0$. To see the competition between the symmetric phase
($W=0$) and the WFH phase ($W=F\neq 0$), we compare the potential
energies at two possible global minima as
\begin{align}
  V_{\text{eff}}[W=0,F] &= a F \pm F^2 + F^3,
\label{eq:poten_s} \\
  V_{\text{eff}}[W=F,F] &= a F + b F^2 + c F^3,
\label{eq:poten_b}
\end{align}
where  $a=a_1'$, $b=\pm 1+b_1'$, and $c=1+b_2'$. 

The gap equation for $F$ is obtained from
$\partial V_{\text{eff}}/\partial F=0$. Eq.~(\ref{eq:poten_s}) has a
global minimum at $F_0>0$, i.e.,
$V_{\text{eff}}[F_0,0] < V_{\text{eff}}[0,0]=0$ is satisfied for $a<0$
(for $+F^2$) and for $a<1/4$ (for $-F^2$). On the other hand,
Eq.~(\ref{eq:poten_b}) has a global minimum at $F_1>0$ either for
($a>0$, $b<0$, $b^2>4ac$) or for $a<0$. Then the condition
$V_{\text{eff}}[F_1,F_1]<V_{\text{eff}}[0,F_0]$ yields an inequality;
\begin{equation}
\begin{split}
 & \biggl\{-(1-3a)^{3/2}+\frac{(b^2-3ac)^{3/2}}{c^2}\biggr\} \\
 & \quad +\biggl(\pm 1-\frac{b^3}{c^2}\biggr)
   -\frac{9}{2}\biggl(\pm a-\frac{ab}{c}\biggr) > 0.
\label{eq:cond-gen}
\end{split}
\end{equation} 

\begin{figure}[htbp]
\includegraphics[width=7cm]{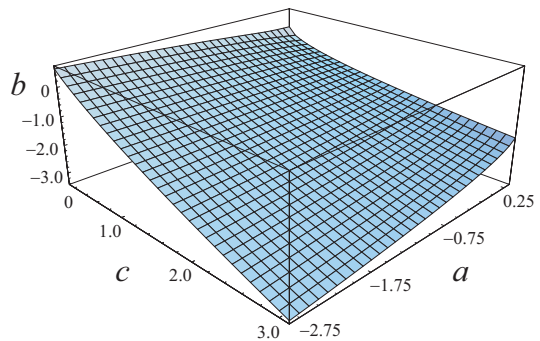}
\caption{In the region below the surface, the WFH phase is favored.}
\label{fig:abc}
\end{figure}

Shown in Fig.~\ref{fig:abc} is the above condition for the negative
$F^2$ term in Eqs.~(\ref{eq:general_fw}) and (\ref{eq:poten_s}). The
region below the surface corresponds to the parameters where the WFH
phase can exist. When the $F^2$ term is positive, it is sufficient to
consider the potential only up to the $F^2$ term in order to 
guarantee the stability of the potential, and  the WFH phase
is realized for $a<0$ and $0<b<1$. The message here is that there is
always a wide parameter region in the Ginzburg-Landau type effective
potential so that the WFH phase is realized.


To study  microscopic mechanism to induce the instability toward
$W \neq 0$ in field theoretical models, let us consider an $O(2)$
$\phi^6$ model in three spatial dimensions with the Hamiltonian
density,
\begin{equation}
 \mathcal{H} = \frac{1}{2} |\nabla \phi|^2
  +\frac{g_2}{2}|\phi|^2 +\frac{g_4}{4}|\phi|^4
  +\frac{g_6}{6}|\phi|^6.
\end{equation}
We assume $g_6>0$ for stability of the system. If the symmetry group
is $O(4)$ instead of $O(2)$, the model is considered as the 3d
effective theory to describe the tricritical behavior of the chiral
phase transition in QCD at finite temperature and baryon density
\cite{step98}.

We treat the model in the Cornwall-Jackiw-Tomboulis (CJT) formalism
\cite{cor74} which is best suited for studying the system with
composite order parameters. The effective action is given in general
by
\begin{equation}
 \Gamma[G,H] = I_0[H]+\frac{1}{2}\mathrm{Tr}\ln G^{-1}+\frac{1}{2}
  \mathrm{Tr}\,G_0^{-1}[H]\, G +\Gamma_2[G,H],
\end{equation}
where $I_0[H]$ and $G_0[H]$ are the tree-level potential and
propagator, respectively. $G$ is the full propagator satisfying
$\delta\Gamma/\delta G=0$. $\Gamma_2$ represents the contributions
from 2 particle irreducible (2PI) diagrams with $G$.

Taking into account the terms up to leading order 2PI diagrams, which
is equivalent to the Hartree-Fock (HF) approximation in many-body
theories, $G$ can be written by the dynamical mass $m_a$ of the field
$\phi_a$. Then the expectation value of the local composite operator
$G_{ab}(x,x)=\langle \phi_a \phi_b \rangle $ reads
\begin{align}
 S_{ab} &\equiv  G_{ab}(x,x)
  =\delta_{ab}\int^\Lambda \frac{\mathrm{d}^3p}{(2\pi)^3}
  \frac{1}{p^2+m_a^2} \notag\\
 &= \delta_{ab}\frac{1}{2\pi^2}\biggl\{\Lambda-|m_a|\arctan\Bigl(
  \frac{\Lambda}{|m_a|}\Bigr)\biggr\},
\label{eq:one-loop}
\end{align}
where $\Lambda$ is the three dimensional cutoff of the loop integral.
The term proportional to $\Lambda$ in Eq.~(\ref{eq:one-loop})
corresponds to the short distance part independent of $m_a$. Thus, the
WFH phase, in which
$\langle\phi_1^2-\phi_2^2\rangle=S_{11}-S_{22}\neq0$ is realized, 
 can 
be characterized by $m_1\neq m_2$ within the present approximation.

\begin{figure}[htbp]
\includegraphics[width=8cm]{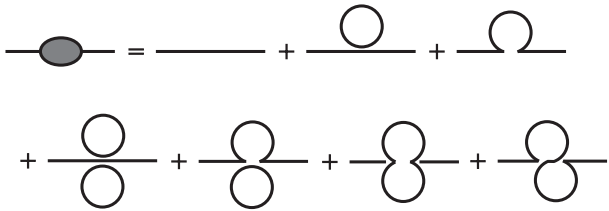}
\caption{An illustration of the gap equation,
 Eq.~(\ref{eq:self-energy}). The Hartree terms correspond to those
 with fully disconnected loops (the second term in the first row and
 the first term in the second row). Other terms with loop(s) are the
 Fock terms.}
\label{fig:mass}
\end{figure}

From now on we take $H=0$ by hand as before to focus on the phase with
$W\neq0$. The dynamical masses, $m_a$, are determined by the gap
equation derived from the variation of the CJT potential or from the
self-consistent HF equation;
\begin{align}
 & \frac{m_a^2}{2} = \frac{g_2}{2}+\frac{g_4}{4}\Bigl(\sum_b S_b+2S_a
  \Bigr) \notag\\
 & \; +\frac{g_6}{6}\biggl\{\Bigl(\sum_b S_b\Bigr)^2+4S_a \sum_b S_b
  +2\sum_b S_b^2+8S_a^2\biggr\},
\label{eq:self-energy}
\end{align}
where $S_a\equiv S_{aa}$ (see Eq.~(\ref{eq:one-loop})). Each term in
Eq.~(\ref{eq:self-energy}) has the graphical representation as shown
in Fig.~\ref{fig:mass}.


The genuine Hartree terms in the right hand side of
Eq.~(\ref{eq:self-energy}) or the fully disconnected graphs in
Fig.~\ref{fig:mass} are index-blind and give equivalent contributions
to $m_1$ and $m_2$. Hence, if we take only Hartree terms, the standard
solution, $m_1=m_2$, is obtained. On the other hand, the Fock terms in
the right hand side of Eq.~(\ref{eq:self-energy}) are connected to the
external lines and depend on the external index $a$. This leads to a
possibility for $m_1\neq m_2$.
  
Of course, the above argument does not guarantee the realization of an
asymmetric solution, because $m_1=m_2$ is also a solution even if we
have Fock terms. To study which solution is more stable, we have
calculated the CJT effective potential written in terms of
$f=|m_1|+|m_2|$ and $w=|m_1|-|m_2|$. For small $m_a$, they are related
to $F=S_{11}+S_{22}$ and $W=S_{11}-S_{22}$ as
\begin{equation}
\begin{split}
 & F=\frac{1}{2\pi^2}\Bigl(2\Lambda-\frac{\pi}{2}f+O(|m_a|^2)\Bigr),\\
 & W=-\frac{1}{2\pi^2}\Bigl(\frac{\pi}{2}w+O(|m_a|^2)\Bigr).
\end{split}
\end{equation} 
For convenience, we use $w$ and $f$ as basic variables in the
following.

Assuming that $m_a$ are small enough compared to $\Lambda$, we expand
the CJT potential in terms of $f$ and $w$ up to the same order with
Eq.~(\ref{eq:general_fw});
\begin{align}
 \bar{V}_{\text{eff}}[w,f]
 & = -\frac{\bar{g}_2}{2} f+\frac{\bar{g}_4}{2}
  f^2+(1-\bar{g}_6)f^3 \notag\\
 &\qquad +\frac{1}{2} \Bigl\{\frac{\bar{g}_4}{2}+3(2-\bar{g}_6)f
  \Bigr\}w^2.
\label{eq:pot2}
\end{align}
Note that $w^2$ terms come from the Fock contributions and the bare
coupling constants and the potential are redefined as
\begin{equation}
\begin{split}
 \bar{g}_6 &= \frac{3}{2\pi^2}g_6, \qquad
 \bar{g}_4 = \frac{24}{\pi}g_4+\frac{36}{\pi^3}g_6, \\
 \bar{g}_2 &= 24g_2-\frac{48}{\pi^2}g_4-
  \frac{144}{\pi^4}g_6, \\
 \bar{V}_{\text{eff}} &= 96\pi V_{\text{eff}}
\end{split}
\label{eq:coupling}
\end{equation}
with the cutoff $\Lambda $ being set to 1. The stability of the system
at large $f$ leads to a condition, $\bar{g}_6 < 1$ (or equivalently
$g_6<2\pi^2/3$ \cite{bar84}). Although the definition of the order
parameter is slightly different from the general analysis in
Eq.~(\ref{eq:general_fw}), obtained effective potential has the same
structure. The correspondence becomes even transparent if we rescale
$\bar{V}_{\text{eff}}$ and $f$ (and $w$) to eliminate two of the
coefficients in Eq.~(\ref{eq:pot2}) to obtain
\begin{equation}
 V_{\text{eff}}[w,f] = a_1 f \pm f^2+f^3
  +\Bigl(\pm \frac{1}{2}+b_2 f \Bigr)w^2,
\label{eq:model-pot}
\end{equation}
where $b_2>3$, which comes from $\bar{g}_6<1$. In the following, we
choose the negative signs in the $f^2$ and $w^2$ terms in
Eq.~(\ref{eq:model-pot}) because the $-\frac{1}{2}w^2$ term, which
originates from the Fock contribution, induces the instability toward
the WFH phase. 
 
\begin{figure}[htbp]
\includegraphics[width=8cm]{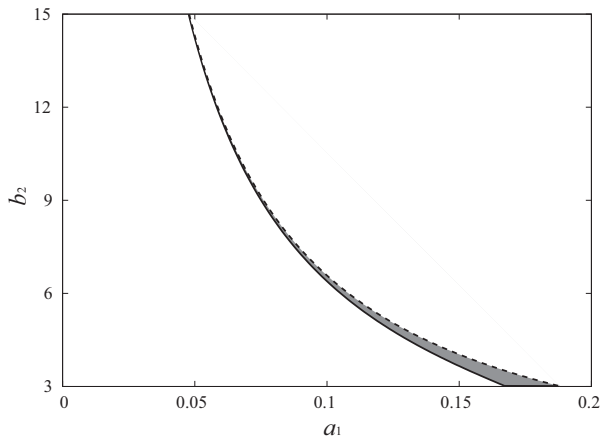}
\caption{Meta-stable WFH phase is allowed in the gray region.}
\label{fig:coupling}
\end{figure}

The condition that $V_{\text{eff}}[w=m_1,f=m_1]$ (the WFH phase) has
at least a local minimum with respect to $m_1$ leads to
$a_1(b_2+1)<3/4$. This is shown by the left hand side of the dotted
curve in Fig.~\ref{fig:coupling}. The stability in the $m_2$ direction
requires $3(b_2+1)+4a_1b_2(b_2-1)+\sqrt{3}\sqrt{3-4a_1(b_2+1)}>0$,
which is shown by the right hand side of the solid curve in
Fig.~\ref{fig:coupling}. As we mentioned, $b_2 >3$ is necessary for
the stability of the potential at large $f$. As a result, the WFH
phase is allowed at least as a local-minimum in the gray region in
Fig.~\ref{fig:coupling}. It can be, however, shown that it is not a
global minimum but a meta-stable phase. In fact, by comparing
Eq.~(\ref{eq:model-pot}) with $f=w$ and Eq.~(\ref{eq:poten_b}), one
finds  $a=a_1$, $b=-3/2$, and $c=1+b_2>4$. This parameter set is too
restrictive and can not satisfy the condition Eq.~(\ref{eq:cond-gen})
which is necessary for the absolute stability. 

Even though it is only
a meta-stable state, an important message here is that the 3d
$O(2)$ $\phi^6$ theory in the HF approximation embodies a driving
force (the Fock term) leading toward the WFH phase. Whether this new
phase is indeed realized as a true ground state or not should be
examined in an approach beyond the HF approximation or by numerical
simulations. In $O(N>2)$ theories, the Fock term is suppressed
relative to the Hartree term for large $N$. Therefore the WFH phase
driven by the Fock term is expected in small $N$ instead of in large
$N$. 


In summary, we have studied higher dimensionful order parameters
associated with the dynamical symmetry breaking which leaves manifest
discrete symmetry. By taking a complex scalar field $\phi$ and its
Ginzburg-Landau potential as an example, we have given a general
criterion to have a novel symmetry breaking pattern $O(2)\to Z_2$
characterized by the order parameter
$\mathrm{Re}\,\langle \phi^2 \rangle \neq 0$. We have demonstrated
that the Hartree-Fock approximation to the 3d $O(2)$ $\phi^6$ theory
gives such a novel phase at least as a meta-stable state. Whether the
new phase exists as a true ground state beyond the HF approximation is
an open question to be studied. We have assumed the stability around
$\langle\phi\rangle=0$ in this Letter; further studies of the
interplay among the three condensates, $\langle\phi\rangle$,
$\langle|\phi|^2\rangle$, and $\langle\phi^2\rangle$ should be also
pursued. The lesson we can learn from our results in this Letter is
that SSB through higher dimensionful order parameters is not so
peculiar and may possibly be realized in more complex systems.
Applications of the idea to e.g.\ gauge field theories and condensed
matter systems are interesting future problems to be studied.


\textbf{Acknowledgments:}
Y.~W.\ is grateful to Ayumu Sugita for his crucial suggestions.
K.~F.\ is supported by Japan Society for the Promotion of Science for
Young Scientists. This work is partially supported by the
Grants-in-Aid of the Japanese Ministry of Education, Science and
Culture (No.~15540254) and the U.S.\ Department of Energy (D.O.E.)
under cooperative research agreement \#DF-FC02-94ER40818.


\end{document}